\documentclass[twocolumn, prA]{revtex4}
\usepackage{graphicx}
\usepackage{amsmath}
\usepackage{stmaryrd}
\usepackage{textcomp}
\usepackage{subfig}

\begin{document}

\title{Interactions Between Rydberg-Dressed Atoms}
\author{J. E. Johnson, S. L. Rolston}
\affiliation{Joint Quantum Institute, Dept. of Physics, University of Maryland and National Institute of Standards and Technology, College Park, MD 20742-4111, U.S.A.}
\date{\today}

\begin{abstract}
We examine interactions between atoms continuously and coherently driven between the ground state and a Rydberg state, producing ``Rydberg-dressed atoms." Because of the large dipolar coupling between two Rydberg atoms, a small admixture of Rydberg character into a ground state can produce an atom with a dipole moment of a few Debye, the appropriate size to observe interesting dipolar physics effects in cold atom systems. We have calculated the interaction energies for atoms that interact via the dipole-dipole interaction and find that due to blockade effects, the $R$-dependent two-atom interaction terms are limited in size, and can be $R$-independent  up until the dipolar energy is equal to the detuning.  This produces $R$-dependent interactions different from the expected $1/R^3$ dipolar form, which have no direct analogy in condensed matter physics, and could lead to new quantum phases in trapped Rydberg systems.
\end{abstract}

\maketitle

\section{Introduction}
	There have been numerous theoretical predictions of novel phases of matter in ultracold atomic systems with long-range interactions, including dipolar crystals \cite{pupillo, Mitra, Zoller, buchler}, supersolids, \cite{dassarma05, goral02, buchler}, striped and checkerboard phases \cite{dassarma05, goral02, kollath08}, and others \cite{Astrakharchik}. These calculations have generally assumed the existence of ground-state polar molecules with dipole moments in the range of $2-5 ea_0$. Although there have been dramatic accomplishments recently in the production of ground-state polar molecules \cite{goral02}, they have yet to be used for such dipolar many-body physics, and are challenging to produce.  As has been pointed out by \cite{pupillo}, Rydberg atoms might also be able to fill this role.  The maximum dipole-dipole interaction between two Rydberg atoms with principle quantum number $n$ is of order $n^4 ea_0$, orders of magnitude larger than needed for predicted dipolar effects, and in fact so large that the interparticle forces would overwhelm any optical trapping forces from an optical lattice, for example. In addition, the typical lifetime of a Rydberg state with $n = 50$ is about $100$ $\mu $s \cite{gallagher}, too short to allow a many-body system to reach equilibrium.   Because the full Rydberg-Rydberg interactions are so strong, we can use a state with only a fraction of those interactions, something we can achieve by creating a wavefunction that is mostly ground state with a small, adjustable Rydberg component, using a coherent coupling, dressing the atom \cite{cohen}.  This would be accomplished with continuous laser  irradiation of ground state atoms, coherently coupling the ground state to the Rydberg state via a one- or two-photon transition.  We can imagine creating atoms that have only 1\% Rydberg character, which is still sufficient to create interesting dipolar physics, and increases the lifetimes to $\sim 10$ ms, where it may be possible for the system to come to equilibrium.  The admixture fraction is controllable by adjusting the coupling laser detuning and intensity, and thus would also give a tunable dipolar coupling between atoms, which could be a useful feature in exploring the effects of long-range interactions.  Additionally, the dipole-dipole interaction could be dependent upon an externally-applied static electric field, either through tuning near a F\"{o}rster resonance \cite{walkerFR}, or by inducing a dipole moment via the Stark Effect \cite{sakurai}.

In what follows, we will assume an idealized Rydberg coupling laser, described solely by a Rabi frequency $\Omega$ and a detuning $\delta$ with respect to the one-atom Rydberg transition.  In practice this would be created through a two-photon excitation with a large intermediate state detuning to assure coherent coupling.  In \cite{pfauecho} they demonstrate coherent coupling between a ground and Rydberg level in Rb with a two-photon Rabi frequency of $\sim 100$ kHz, and an intermediate state detuning of $\sim 500$ MHz. Our goal is to create Rydberg-dressed atoms with a wavefunction 
	
\begin{equation}
|\psi \rangle = \alpha |g \rangle + \beta |r \rangle
\end{equation}
where $|g \rangle$ is the ground state and $|r \rangle$ is the Rydberg state.  Such a state would have a spontaneous decay rate of 
\begin{equation}
\gamma \sim | \langle g | \vec{d} | \psi \rangle |^2 \sim \beta^2 \gamma_r
\end{equation}
where $\gamma_r $ is the Rydberg decay rate, and $\vec{d}$ is the dipole operator for spontaneous emission from the Rydberg state \cite{sakurai}. It is tempting to then simply calculate the dipole-dipole interaction between the two dressed states as
\begin{equation}
\epsilon_{int}= \langle\psi | U_{dd} |\psi \rangle  = \beta^2 \langle r| U_{dd} | r \rangle  = \beta^2 \epsilon_r
\end{equation}
where $U_{dd}$ is the usual dipole-dipole operator between Rydberg states, and $\epsilon_r$ is the full interaction energy between two Rydbergs, which can be of order 10 GHz for R = 1 $ \mu$m and n $\sim 50$.  As we will see below, this expression is in general invalid, because the atom-atom interactions will cause a blockade effect \cite{jaksch98, walker, grangier}, such that the two-atom wavefunction contains much less than $\beta^2$ of the $|r\rangle |r\rangle$ state.  The correct procedure is to calculate the dressed states for two atoms simultaneously.  The blockade effect will arise naturally out of the dressed eigenstates of this two-atom system.

\begin{figure}[t]
\centering
\includegraphics[width=80mm]{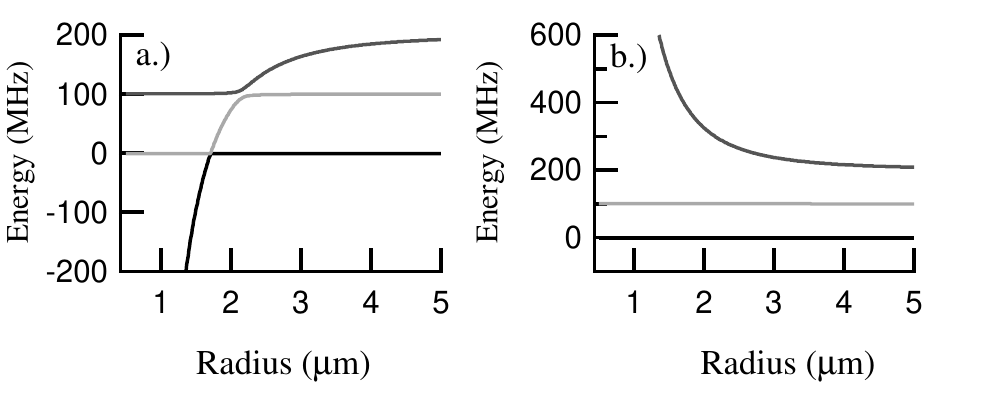}
\caption{The energy levels of the two-atom state for $\Omega /2\pi = 10$ MHz, $\delta/2\pi = 100$ MHz, and a.) $c_3/2\pi= -1000$ MHz$\times \mu m^3$, and b.)  $c_3/2\pi= 1000$ MHz$\times \mu m^3$.}
\label{fig:2atomenergy}
\end{figure}

\begin{figure}[b]
\centering
\includegraphics[width=80mm]{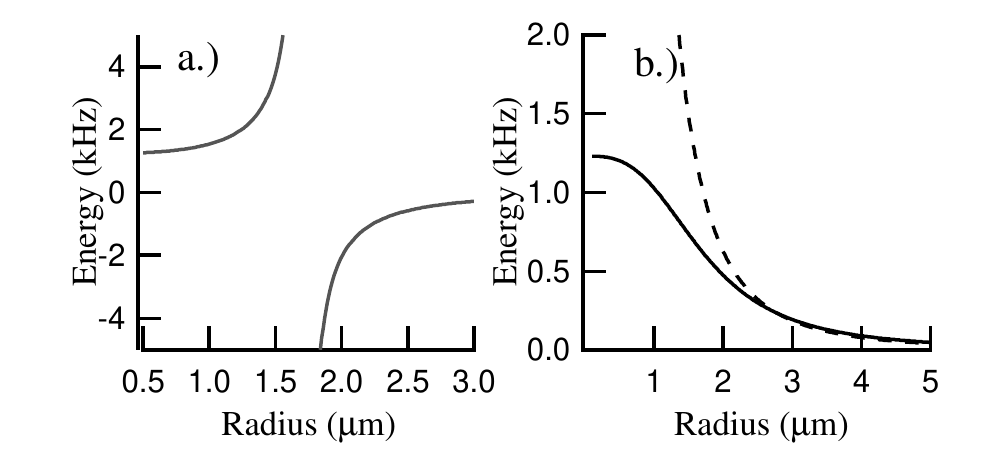}
\caption{A detail of the interaction energies for the cases shown in Fig. \ref{fig:2atomenergy}.  Part b.) shows the convergence with an interaction $\sim 1/r^3$ at longer distances.}
\label{fig:2atomblowup}
\end{figure}

\section{Two Interacting Dressed Atoms}
	The simplest case of interacting dressed atoms to consider is the two-atom case.  We could write a $4\times 4$ Hamiltonian matrix in the basis $|gg\rangle,|gr\rangle,|rg\rangle,|rr\rangle$, but if instead we use a basis with $1/\sqrt{2}(|gr\rangle \pm |rg\rangle)$, the antisymmetric state is uncoupled, and can be ignored.  The resulting $3\times 3$ Hamiltonian in the basis  $|gg\rangle, 1/\sqrt{2}(|gr\rangle + |rg\rangle),|rr\rangle$ can be written as
\begin{equation}
H= \hbar\left(\begin{array}{ccc}0 & \frac{\Omega}{\sqrt{2}} & 0 \\\frac{\Omega}{\sqrt{2}} & \delta & \frac{\Omega}{\sqrt{2}} \\ 0  & \frac{\Omega}{\sqrt{2}} & 2\delta +U_{dd} \end{array}\right).
\end{equation}	

If $U_{dd}$ = 0, we recover the dressed states and eigenenergies for two independent dressed atoms, as expected.    After diagonalizing the matrix, we find the energy levels as shown in Fig. \ref{fig:2atomenergy}, where we have used a simple $c_3/R^3$ for the dipole-dipole interaction energy, with $c_3/2\pi = \pm 1$ GHz$\times\mu$m$^3$, typical for a dipole-dipole interaction for $n \sim 40-50$, assuming an applied static electric field, for example. We subtract off the constant light shift due to the coupling laser so that we plot the interaction energy solely due to the two-atom effects (we also ignore the angular dependence of the dipole interaction, effectively assuming a fixed interatomic axis).  Figure \ref{fig:2atomenergy} shows two cases, for positive and negative $c_3$.  In the case of negative $c_3$ there is an avoided curve crossing when the laser is two-photon resonant with the dipole-shifted $|rr\rangle$ state, i.e. $2\delta=-c_3/R^3$. We consider both the positive and negative $c_3$ cases with the detuning chosen so that the interaction energy at short range is positive.  Because the trapping potential confines the atomic gas, an attractive interaction would lead to collapse of the cloud rather than emergent order \cite{koch08}.  Figure \ref{fig:2atomblowup} shows a detail of the interaction energies. Note that inside this avoided crossing the eigenenergy of the state that connects to the ground state (the state of interest in the context of a Rydberg dressed atom) is almost independent of $R$, while outside the crossing it falls off rapidly with $R$.   For positive $c_3$ (Figs. \ref{fig:2atomenergy}b, \ref{fig:2atomblowup}b) there is no avoided crossing, yet the eigenenergy also becomes independent of $R$ at distances less than about $0.5$ $\mu m$, which is comparable to small optical lattice spacings.  This $R$-independence is a consequence of the blockade phenomenon.  The large $U_{dd}$ term in the Hamiltonian makes the coupling from the single-atom excited state to the doubly excited state off-resonant, significantly reducing the $|rr\rangle$ component of the two-atom wavefunction (to much less than $\beta^2$).


\section{Two-Atom Eigenstates}
	We can find the eigenstates of this matrix, as a function of $R$.  Examination of these eigenvectors shows that for the eigenvector associated with the lowest energy eigenvalue, the system is predominantly in the ground state, with small amounts of the two excited states mixed in.  Using these state populations, we can calculate the percentage of Rydberg in the admixed state from the expression
\begin{equation}
2|\langle rr| \psi \rangle |^2+\frac{1}{2}|\langle gr|\psi\rangle +\langle rg|\psi\rangle |^2
\end{equation}
which shows that, for the values used previously, the state is $0.5\%$ in the Rydberg state, which yields a lifetime of about $25$ ms.  Inside the blockade radius, the system cannot be in the state where both atoms are excited to the Rydberg level, due to blockade, so the Rydberg characteristic inside the blockade radius comes from the eigenstate where only one of the atoms is excited to the Rydberg level.

	In the limit when $U_{dd} \gg \delta$, at small $R$ we can expand the ground-state energy of this matrix near $R=0$. The ground state energy inside the blockade radius becomes $E_{gg}/\hbar\approx (1/2)(\delta-\sqrt{2\Omega^2+\delta^2})$.  Taking the difference between this energy and the non-interacting ground state energy gives an energy due to interactions of $E_{int}/\hbar \sim (1/8)(\Omega^4/\delta^3)$, for $\delta \gg \Omega$. For experiments with cold atoms, interaction energies need to be of the order of kHz, of the same order as ground state atom-atom interactions for atoms trapped in a single lattice site (such as is relevant for the Bose-Hubbard Hamiltonian).  From this expression, we can see that it is possible to achieve such magnitude of interaction energy with Rabi frequencies of order 10 MHz and detunings of order 100 MHz, while at the same time keeping the Rydberg fraction to 1\% or less (necessary for sufficiently small spontaneous emission rates).  The $\Omega^4$ dependence of the interaction energy puts a premium on large Rabi frequencies, which will be the primary experimental challenge, especially if it needs to be implemented over a large spatial volume.

A second limiting case we can consider is when the atoms are separated by a distance much greater than the blockade radius, or $2\delta \gg U_{dd}$.   We find that the ground-state energy eigenvalue dependent on $R$ scales as $U_{dd}(1/16)\Omega^4/\delta^4$.  This is as expected:  if we can ignore blockade, the admixture ratio of the excited state for a single atom goes as $\beta\sim (1/2)\Omega/\delta$, so the doubly excited state mixture would go as $\beta^2$, and the matrix element $\langle rr |V_{dd}| rr \rangle$ scales with $\beta^4$,  as found. Note that once again the interaction energy scales as $\Omega^4$.

\begin{figure}[t]
\centering
\vspace{-0.35in}
\includegraphics[width=60mm]{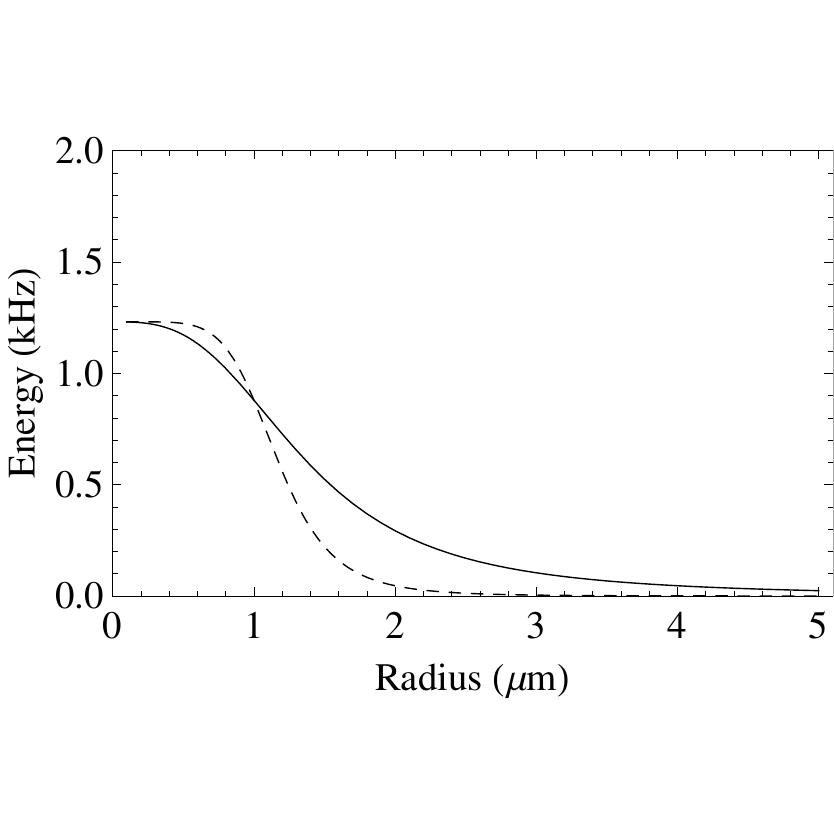}
\vspace{-0.35in}
\caption{The energy of the dressed state, for two different interactions: the dipole-dipole that varies as $1/R^3$ (solid), and the Van der Waals that varies as $1/R^6$ (dashed), for $\delta/2\pi = 100$ MHz, $c_3/2\pi= 1000$ MHz$\times \mu m^3$, and $c_6/2\pi= 500$ MHz$\times \mu m^6$.}
\label{fig:energyrange}
\end{figure}

The above discussions are independent of the actual nature of the atom-atom interaction.  Although presented in terms of a pure dipole-dipole interaction,  $\sim c_3/R^{3}$, the same blockade mechanism applies for any interaction larger than the detuning.  Figure \ref{fig:energyrange} plots the energy of the dressed state connected to the ground state for two interactions : pure dipole-dipole ($c_3/R^{3}$) and pure Van der Waals ($c_6/R^{6}$).  A potential such as that shown in \cite{walkerFR} for two interacting 50s Rb atoms that is dipolar at short range and van der Waals at long range would fall in between the curves shown in Fig. \ref{fig:energyrange}.  In all cases the eigenenergies evolve similarly with $R$, with the only difference the rate of change with $R$.

\section{Possible Experimental Implications}
The $R$ dependence of the interactions shown in Figs. \ref{fig:2atomenergy} and \ref{fig:2atomblowup} is unlike the dipole-dipole $1/R^3$ interaction assumed in most theoretical treatments examining the many-body physics effects of these interactions.  It is possible to recover an interaction of this form by working at interatomic distances that are large compared to the critical distance set by the condition where $2 \delta = U(R_c)$, but it will be challenging to work in this regime with reasonable parameters.  There are a number of requirements: the size of the interaction energy should be sufficiently large, in the 1-10 kHz range (similar to a BEC chemical potential or the atom-atom interaction in a typical lattice); the spontaneous emission lifetime should be sufficiently long such that the many-body system can come into equilibrium ($\geq 10$ ms); the relevant interatomic distance should match typical distances, either a lattice spacing or the average interparticle distance in a gas (0.5 - 1 $\mu$m). It is difficult to achieve all these requirements  for reasonable values of the coupling Rabi frequency.  Ref \cite{pupillo} suggests working at rather low principle quantum numbers (n=20), but requires a large Rabi frequency (100 MHz) and yields a short lifetime (5 ms). 

Producing standard dipolar physics with Rydberg-dressed atoms seems challenging at best.  Another approach is to look at the $R$-dependent interactions that are possible with accessible experimental parameters and consider new many-body physics that is not possible for normal condensed matter systems.  While such many-body states are beyond the scope of this work, we can speculate.  If we consider the simpler case without any level crossings (Fig. \ref{fig:energyrange}), we have an interaction that is repulsive and $R$-independent up to a critical radius, where it falls off rapidly towards zero.  We can consider the impact of this interaction on a gas and on atoms trapped in an optical lattice.  If the density was low enough for a gas of atoms, this interaction would look essentially like a contact interaction.  At very high densities where the interatomic spacing is small compared to $R_c$ the interactions just provide an overall energy shift, presumably with little impact on the system.  In an intermediate density region, however, the cutoff at $R_c$ can be expected to have significant impact on the correlation functions of the gas, perhaps leading to exotic phases and order in the cloud.  For an optical lattice, the addition of a nearest neighbor interaction to the Bose-Hubbard model leads to such phenomena as a supersolid phase \cite{dassarma05}.  With the Rydberg-dressed atom interactions we could have a situation where the first $n$-neighbor interactions are the same, followed by little or no interactions.  As this has no condensed matter analog, prediction is difficult, but novel phases seem likely to exist.  Finally, because parameters exist that achieve reasonable interaction strengths and lifetimes with detunings $\leq$ 1 GHz, we can also consider introducing spin dependence into such interactions.  If a two-level spin system is established between two sublevels in the different hyperfine states of an alkali atom such as Rb, we can map all these interactions onto a spin-dependent model, as only one of the levels would interact with the dressing laser.  

\begin{figure}[b]
\centering
\vspace{-0.35in}
\includegraphics[width=60mm]{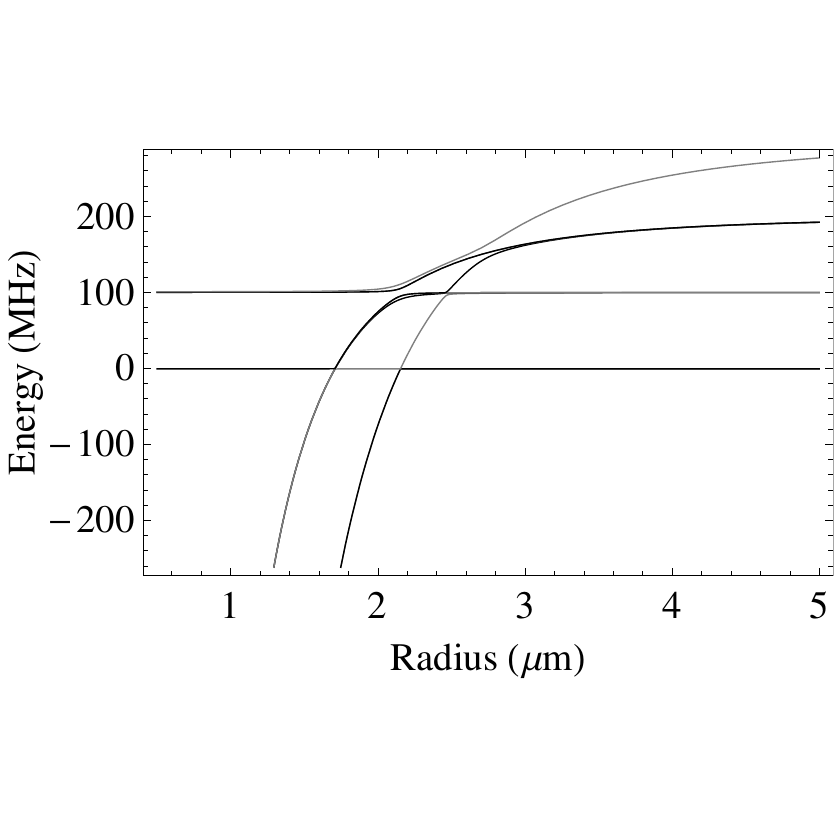}
\vspace{-0.35in}
\caption{The energy levels of the three-atom system in the two-dimensional configuration for $\Omega /2\pi = 10$ MHz, $\delta/2\pi = 100$ MHz, and $c_3/2\pi= -1000$ MHz$\times \mu m^3$.}
\label{fig:3atomtriangle}
\end{figure}

	One important and simplifying assumption we have made with this treatment is ignoring any short range physics, i.e. assuming a single dipolar coupling at small $R$.  This assumption will not be valid when the dipole-dipole shift is on the order of the spacing between neighboring levels, and the energy spectrum becomes extremely crowded and complicated at short range (see \cite{shaffer} for example).  More detailed calculations beyond simple dispersion expansions of the dipole coupling are needed to understand the energy spectrum as well as potential loss mechanisms.  The effect of the dressing at short distance may be more robust than it seems, however, because it depends solely on the blockade effect.  As long as the interactions dominate the detuning their actual value is irrelevant.  Of course if the coupling laser couples directly to a short-range molecular state, potentially large losses would ensue.

\begin{figure}[t]
\centering
\vspace{-0.35in}
\includegraphics[width=60mm]{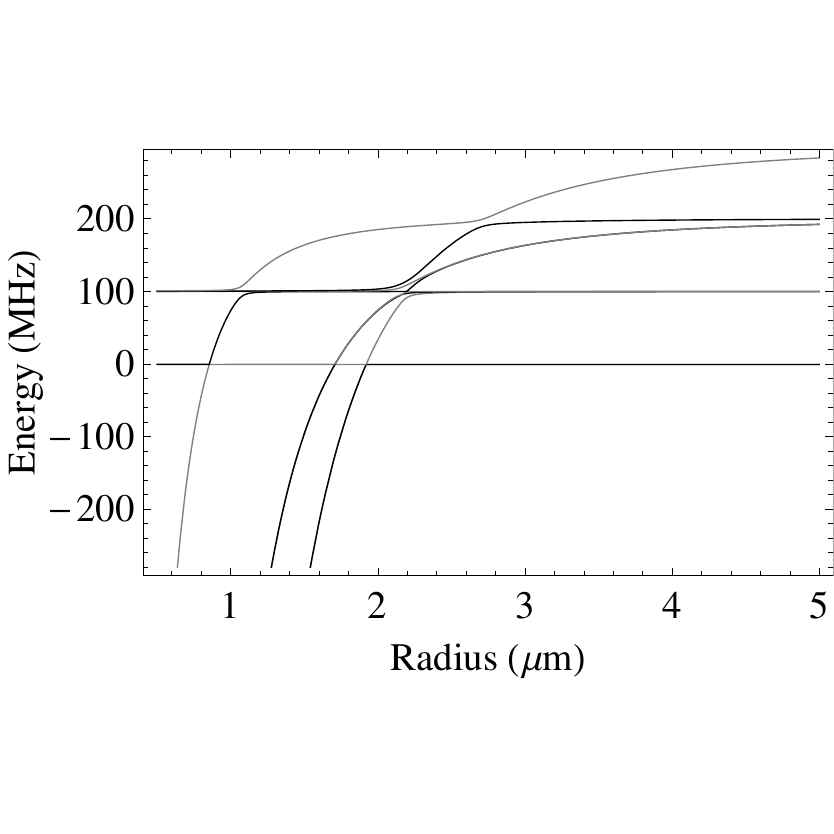}
\vspace{-0.35in}
\caption{The energy levels of the three-atom system in the one-dimensional configuration for $\Omega /2\pi = 10$ MHz, $\delta/2\pi = 100$ MHz, and $c_3/2\pi= -1000$ MHz$\times \mu m^3$.}
\label{fig:3atomline}
\end{figure}

\section{Three Interacting Dressed Atoms}
	So far, we have examined only the simplest interacting system: two interacting dressed atoms.  In order to extend these calculations to many-body systems in the future, it is necessary to explore a more complicated system of atoms.  The next level of complextiy would be a system of three interacting dressed atoms.  We can examine the eigenvalues for a system of three dressed atoms, where the Hamiltonian has multiple independent interaction terms on the diagonal, in two simple configurations: a one-dimensional configuration that examines next-nearest neighbor interactions, and a two-dimensional configuration.  The one-dimensional configuration is a line of the three atoms, with distance $R$ between adjacent atoms, and the two-dimensional configuration is each atom on the vertex of an equilateral triangle with side $R$. 

	The Hamiltonian of atoms arranged in the two-dimensional configuration has two independent diagonal terms.  Because the atoms are equidistant, the diagonal terms for each amount of excitation are degenerate and we can reduce the $8\times 8$ matrix to a $4\times 4$ matrix

\begin{equation}
H= \hbar\left(\begin{array}{cccc}0 & \frac{\Omega}{\sqrt{2}} & 0 & 0 \\ \frac{\Omega}{\sqrt{2}} & \delta & \sqrt{\frac{3}{2}}\Omega & 0 \\ 0 & \sqrt{\frac{3}{2}}\Omega & 2\delta+\frac{c_3}{R^3} & \Omega \\ 0 & 0 & \Omega & 3\delta +3\frac{c_3}{R^3} \end{array}\right)
\end{equation}

using the basis: $|ggg\rangle, 1/\sqrt{3}(|rgg\rangle+|grg\rangle+|ggr\rangle, 1/\sqrt{3}(|rrg\rangle+|grr\rangle+|rgr\rangle, and |rrr\rangle$.  In this case, there are avoided crossings for $2\delta=-c_3/R^3$, as well as for $3\delta=-3c_3/R^3$ for negative detunings and attractive interactions. For a given detuning, the second of these crossings occurs at a larger interparticle spacing, and is a narrower avoided crossing.  This means that the larger interaction occurs at the first, closer, radius.  Figure \ref{fig:3atomtriangle} shows the energy levels of this system.  For negative $c_3$, the interaction energy looks similar to the two-atom case, with an enhanced value for the same Rabi frequency and detuning.

\begin{figure}[b]
\centering
\includegraphics[width=80mm]{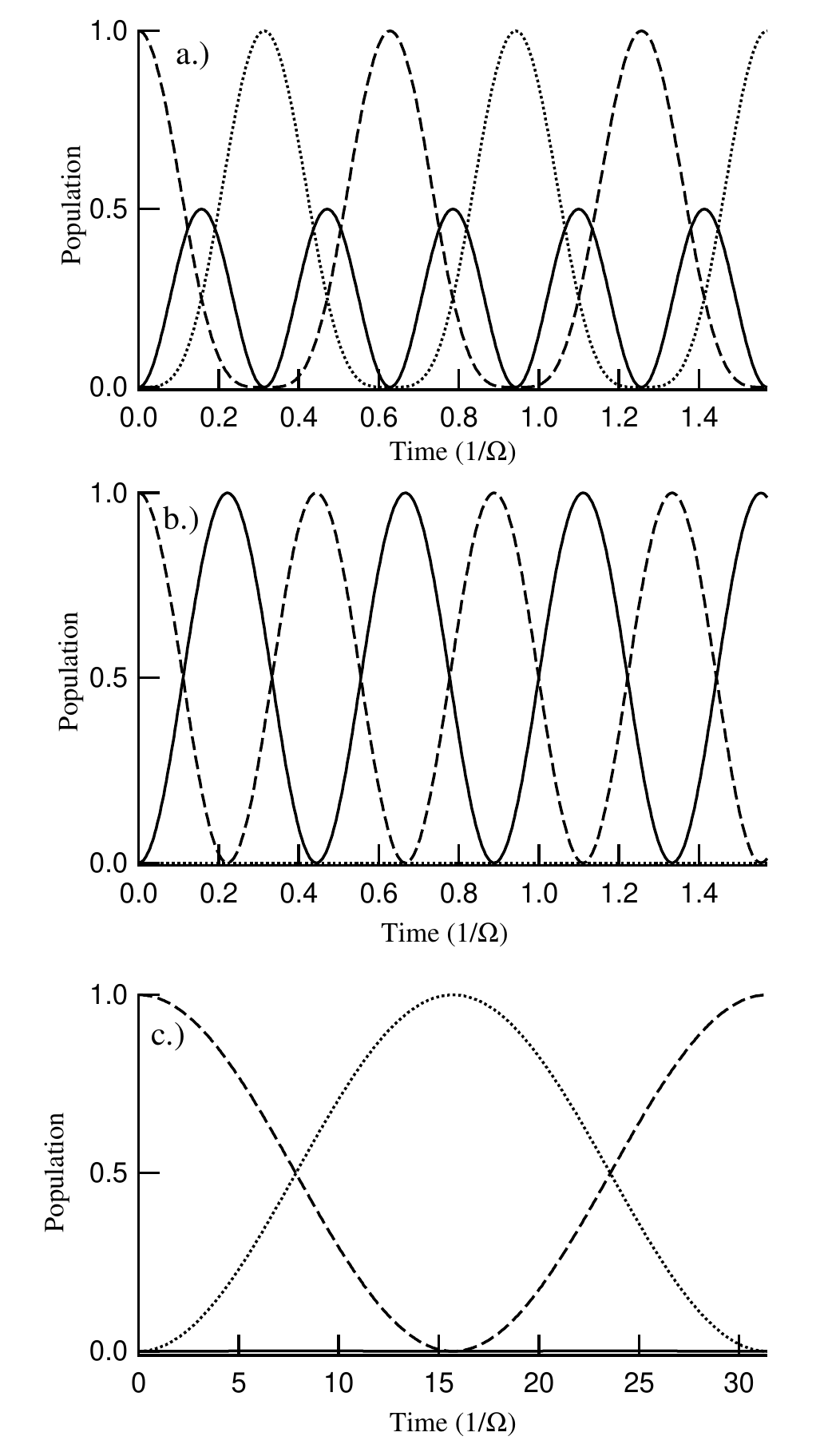}
\caption{The time-dependent evolution of the three eigenstates for atoms separated by $1$ $\mu$m for a.) no atom-atom interactions, b.) non-zero interations, but no detuning, and c.) both detuning and interactions.  The colors show the evolution of the states $|gg\rangle$(dashed), $1/\sqrt{2}(|gr\rangle+|rg\rangle )$(solid), and $|rr\rangle$(dotted).}
\label{fig:timedependence}
\end{figure}

	In the one-dimensional case, where the three atoms lie on a line, we find an additional avoided crossing, where $2\delta=-c_3/(8R^3)$.  This crossing is at an even closer radius, and is wider than the other two. Figure \ref{fig:3atomline} shows these avoided crossings as part of the energy level plot.  In this reduced-dimension case, we see that the addition of a next nearest neighbor breaks the degeneracy among some of the diagonal terms, requiring the full $8\times 8$ matrix for evaluation.  Further expanding these calculations to consider a system of four atoms leads only to additional avoided crossings at longer ranges in the energy spectrum.

\section{Time-Dependent Effects}
	Although up until this point we have been concerned with the steady-state dressed atoms, since we have expressions for the eigenvalues, it is simple to calculate the time dependence for a sudden turn-on of the laser fields.  This can be done by projecting the ground state of the bare atom onto the dressed atom basis, evolving the dressed-atom states in time with the phase factor $\exp(-i \epsilon_i t)$ and then projecting back to the bare-atom basis.  Figure \ref{fig:timedependence} shows three cases: the evolution of the states $|gg\rangle$, $1/\sqrt{2}(|gr\rangle+|rg\rangle )$, and $|rr\rangle$ for a.) no atom-atom interactions, b.) atom-atom interactions with $\delta$ = 0 (single-atom resonance), and c.) atom-atom interactions on the two-atom resonance ($2 \delta = U(R)$).  For case a.) we see Rabi flopping of two independent atoms, with all the population in $|rr\rangle$ for a $\pi$-pulse, and population in all three states during the rest of the evolution.  For case b.) the blockade effect is apparent, in that there is no observable excitation of the doubly excited state, and the system Rabi flops between $|gg\rangle$ and  $|gr\rangle+|rg\rangle$.  For case c.) there is no population of single Rydberg atoms and the system flops between $|gg\rangle$ and $|rr\rangle$.  Note the time scale is longer, because this is a two-photon coupling with a correspondingly smaller Rabi frequency.  

\begin{figure}[t]
\centering
\vspace{-0.35in}
\includegraphics[width=60mm]{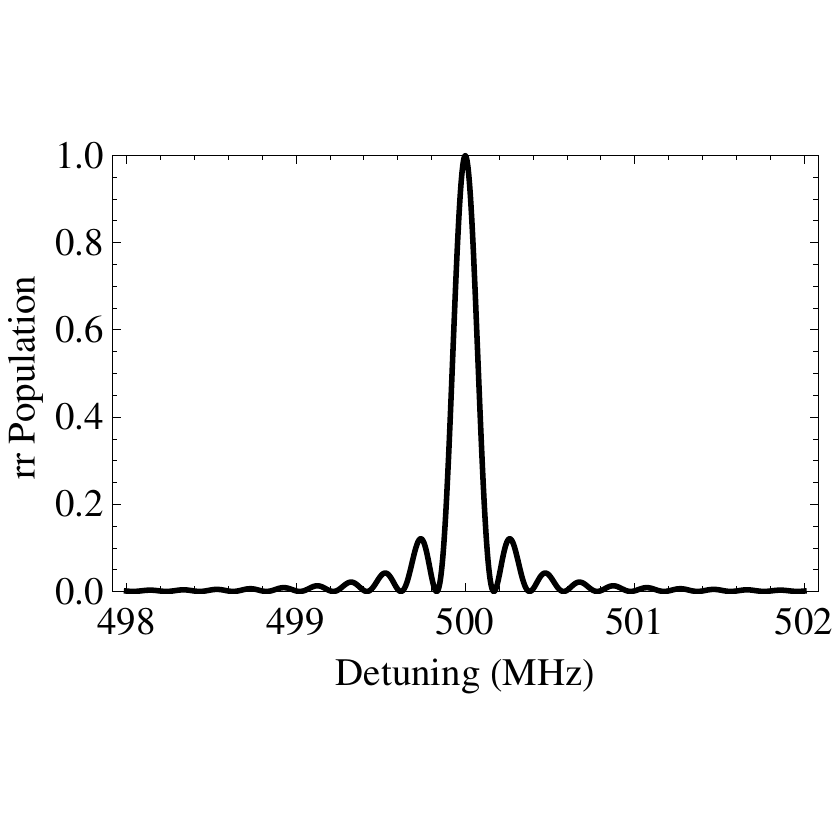}
\vspace{-0.35in}
\caption{The population of the doubly-excited state after a pulse of duration $t=16/\Omega$ at different detunings.}
\label{fig:popvdet}
\end{figure}

	As can be seen in the figure, using a $\pi/2$-pulse with respect to the two-photon Rabi frequency should be a simple way to create an entangled state, $|gg\rangle +|rr\rangle$. Only atom pairs that are at the correct distance to be two-photon resonant will Rabi flop, so this should allow the creation of entanglement of pairs of atoms with ground-state atoms in between, for example.  Figure \ref{fig:popvdet} shows the population at the end of a pulse of duration $t=16/\Omega$ as a function of the detuning.  For small detunings the blockade is evident as there is practically no population of the state $|rr\rangle$.  At the two-photon resonance condition, all of the population can get excited into $|rr\rangle$ for the appropriate pulse duration.

\section{Conclusion}

In summary we have considered creating Rydberg-dressed atoms, with small admixtures of Rydberg character, to study dipolar many-body physics.  We find that the blockade phenomena prevents large interaction energies at small distances, set by a critical radius where the detuning is equal to the dipolar interaction energy.  The resulting $R$-dependent interaction energy can have sufficient size to be relevant for many-body physics of cold atoms, but its functional form is fundamentally different from the $1/R^3$ dipolar interaction that one expects.  Instead it tends to a constant value at small $R$ with a crossover to $1/R^3$ at the critical distance.  This form is unlike any interactions in condensed matter physics and may open up the possibility of novel many-body states.  While these calculations point to experimental parameters that are challenging to achieve, in the future we hope to implement an experiment based on these calculations.  A recent paper, \cite{pfaudressed}, discusses further many-body effects of the interactions of dressed Rydberg atoms.

We acknowledge discussions with J. Porto, S. Das Sarma, and G. Pupillo.  This work is partially supported by the JQI NSF-PFC.

\end{document}